\documentclass[aps,pre,10pt,twocolumn,tightenlines,superscriptaddress,amsmath,amssymb,final,letterpaper,notitlepage,nofootinbib]{revtex4-1}
\usepackage{graphicx}
\usepackage{subfigure}
\usepackage{epstopdf}
\usepackage{times}
\usepackage{bm}
\usepackage[dvipsnames]{xcolor}
\usepackage{tcolorbox}
\usepackage{hyperref}
\usepackage{mathtools}
\usepackage{amsmath}

\graphicspath{{figs/}}

\begin{document}
\title{Source-sink cooperation dynamics constrain institutional evolution in a group-structured society}



\author{Laurent \surname{H\'ebert-Dufresne}}
\thanks{These two authors contributed equally}
\affiliation{Department of Computer Science, University of Vermont, Burlington VT, USA}
\affiliation{Vermont Complex Systems Center, University of Vermont, Burlington VT, USA}
\affiliation{D\'{e}partement de physique, de g\'{e}nie physique et d'optique, Universit\'{e} Laval, Qu\'ebec (Qu\'ebec), Canada, G1V 0A6}

\author{Timothy M. \surname{Waring}}
\thanks{These two authors contributed equally}
\affiliation{School of Economics, University of Maine, Orono ME, USA}
\affiliation{Mitchell Center for Sustainability Solutions, University of Maine, Orono ME, USA}

\author{Guillaume \surname{St-Onge}}
\affiliation{D\'{e}partement de physique, de g\'{e}nie physique et d'optique, Universit\'{e} Laval, Qu\'ebec (Qu\'ebec), Canada, G1V 0A6}
\affiliation{Centre interdisciplinaire en mod\'{e}lisation math\'{e}matique, Universit\'{e} Laval, Qu\'ebec (Qu\'ebec), Canada, G1V 0A6}

\author{Meredith T. \surname{Niles}}
\affiliation{Department of Nutrition and Food Sciences, University of Vermont, Burlington VT, USA}

\author{Laura Kati \surname{Corlew}}
\affiliation{Department of Social Science, University of Maine at Augusta, Bangor ME, USA}

\author{Matthew P. \surname{Dube}}
\affiliation{Department of Computer Information Systems, University of Maine at Augusta, Bangor ME, USA}

\author{Stephanie J. \surname{Miller}}
\affiliation{School of Biology and Ecology, University of Maine, Orono ME, USA}

\author{Nicholas \surname{Gotelli}}
\affiliation{Department of Biology, University of Vermont, Burlington VT, USA}

\author{Brian J. \surname{McGill}}
\affiliation{School of Biology and Ecology, University of Maine, Orono ME, USA}
\affiliation{Mitchell Center for Sustainability Solutions, University of Maine, Orono ME, USA}


\begin{abstract} 
Societies change through time, entailing changes in behaviors and institutions. We ask how social change occurs when behaviors and institutions are interdependent. We model a group-structured society in which the transmission of individual behavior occurs in parallel with the selection of group-level institutions. We consider a cooperative behavior that generates collective benefits for groups but does not spread between individuals on its own. Groups exhibit institutions that increase the diffusion of the behavior within the group, but also incur a group cost. Groups adopt institutions in proportion to their fitness. Finally, cooperative behavior may also spread globally. As expected, we find that cooperation and institutions are mutually reinforcing. But the model also generates behavioral source-sink dynamics when cooperation generated in institutional groups spreads to non-institutional groups, boosting their fitness. Consequently, the global diffusion of cooperation creates a pattern of institutional free-riding that limits the evolution of group-beneficial institutions. Our model suggests that, in a group-structured society, large-scale change in behavior and institutions (i.e. social change) can be best achieved when the two remain correlated, such as through the spread successful pilot programs. 
\newline
\newline
\textbf{Keywords:} cooperation, source-sink dynamics, institutions, behavioral diffusion, cultural evolution 

\end{abstract}

\maketitle

\section{Introduction}
There is broad agreement that global crises such as climate change and pandemic response will require society to change. Therefore, a robust science of social change is needed. It has recently been suggested that the study of collective behavior should be considered a crisis discipline \cite{Bak-Coleman_2021}. Recent research, including the study of social tipping points \cite{Andreoni_2021}, their impact on policy efficacy \cite{Efferson_Vogt_Fehr_2020, Efferson_Vogt_2018}, and the feedbacks between policy and preferences \cite{Schmelz_2021} suggests a science of social change may be possible. But what constitutes social change? At a minimum, we suggest that social change must include both change in individual behaviors and change in institutions. Here we study a model of social change that couples behavioral change with policy evolution.

The most difficult societal problems require both widespread change in individual behavior and institutional change. But individual behavior and group-level institutions are mutually interdependent, making solutions difficult to describe and achieve. On the one hand, large-scale behavioral change is slow or unpredictable unless supported by group-level institutions and policy. On the other hand, policy changes often follow from broad changes in individual preferences, beliefs or behavior (e.g. gay marriage, civil rights). For example, no serious proposals for addressing climate change rely on the spread of voluntary individual behavior alone; institutional support such as requirements and enforcement are clearly required. Yet without widespread individual support for climate policy, nations are unlikely to make institutional and structural change. This reciprocal causation makes addressing major societal goals particularly challenging because there the two most common approaches focus on only one aspect of the system. A policy-first approach uses a top-down change in institutions to affect large-scale behavioral change, for example by requiring and enforcing a behavior (e.g. speeding fines, tax incentives, smoking bans). Alternatively a behavior-first approach seeks to encourage voluntary behavioral change through bottom-up mechanisms such as unenforced behavioral standards (e.g. social distancing guidelines, voluntary participation in environmental programs, and public service announcements generally). Both approaches have benefits and drawbacks. 

\begin{figure}
    \centering
    \includegraphics[width=\linewidth]{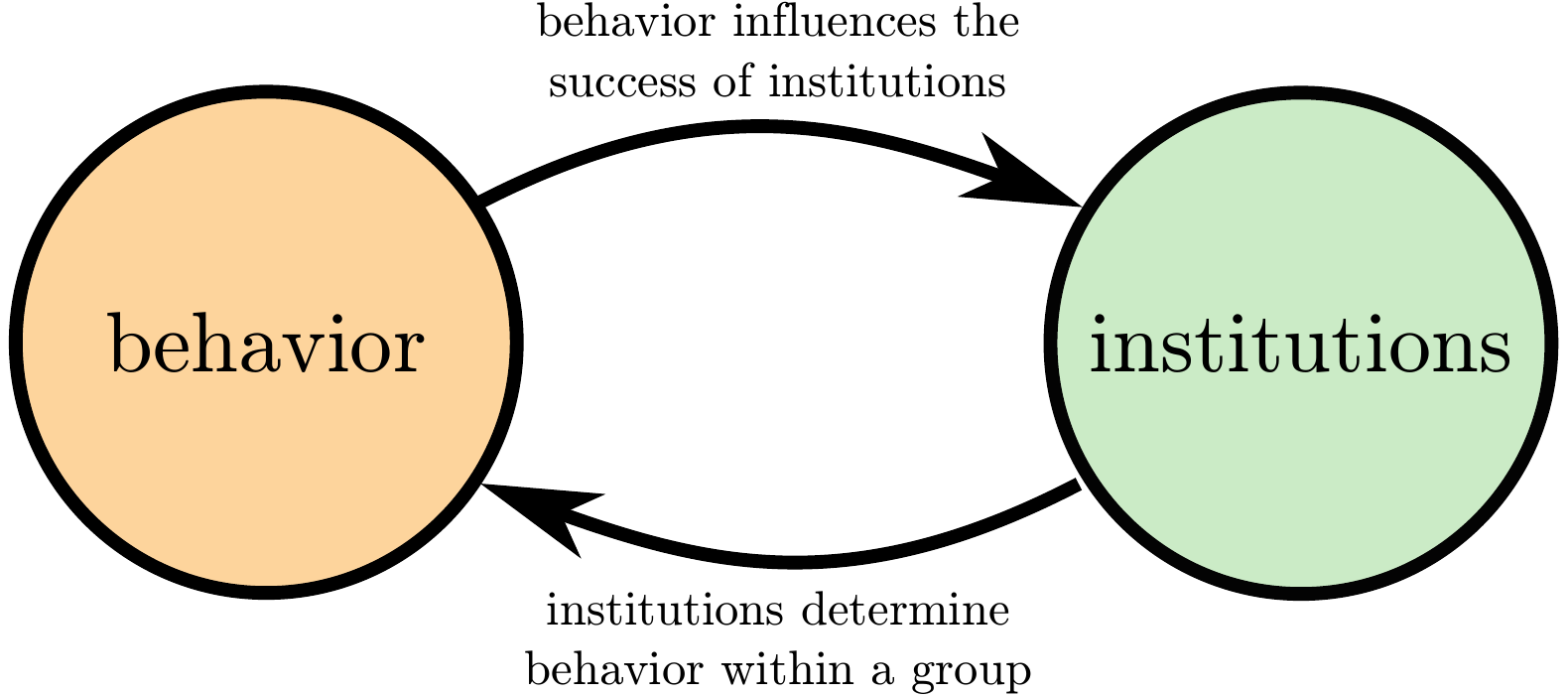}
    \caption{\textbf{Individual behavior and group institutions form a dynamical system with reciprocal causation.} To understand the endogenous nature of social change we must model the causal feedbacks between institutions and behavior, and to influence social change we must be able to find and move toward beneficial states that are self-reinforcing.}
    \label{fig:feedback}
\end{figure}

Policy analysis is complex and policy outcomes are heavily determined by contextual factors. Policy design must therefore focus on local impacts, and thus often ignores where institutions come from, both in terms of how they originate via the policy process \cite{sabatier_2014}, and in terms of how policies spread via policy diffusion \cite{shipan_2012, gilardi_2010}. Institutions, governance, laws, and rules emerge from the complex inter-individual process of cultural evolution and self organization within human groups \cite{Hodgson_2002, Muthukrishna_2020}. As a result, institutions are themselves distributions of individual cultural traits and behaviors \cite{Smaldino_2014}. However, research on the efficacy of policy options e.g. \cite{Bailey_2011, McDonald_2006, Nelson_2013} tends to be isolated from research on policy diffusion \cite{Tews_2003, shipan_2012}, and from research on the evolution of institutions \cite{Hodgson_2002, Lewis_2012, Richerson_2001, Bowles_2003}. Research that does link behavioral change to policy change typically focuses on only one half of the causal system. For example, a policy analyst might study how new COVID-19 policies determined individual mask-wearing behavior. Conversely, a social scientist might study the spread of vaping behavior through a social network and its impact on institutional responses. These are useful and necessary approaches; but behavior and institutions are strongly reciprocally linked, as summarized in Fig.~\ref{fig:feedback}.

By contrast, a behavior-first approach relies on the endogenous processes of behavioral adoption and transmission. Behavior-first strategies such as nudges \cite{sunstein_2014} and viral advertising efforts, seek to leverage endogenous behavioral change for certain outcomes. Research on social tipping points shows that social momentum can be harnessed to improve  well-being by boosting the spread of beneficial social norms and behaviors \cite{Andreoni_2021}. Similar research shows that these social tipping points can be caused by conformity and behavioral transmission and can create behavioral spillovers that determine the efficacy of policy \cite{Efferson_Vogt_2018, Efferson_Vogt_Fehr_2020, Muthukrishna_2020}. When successful, behavior-first approaches such as voluntary guidelines can lead to the diffusion of behavior within communities. Research of this sort typically seeks to find a social tipping point past which the beneficial behavior becomes ubiquitous \cite{Berger_2021}. If the behavior becomes associated with values or social identities \cite{smaldino_adoption_2017}, the result can be the evolution of a self-reinforcing system of social norms: a new standard of behavior \cite{Schmelz_Bowles_2021, Schmelz_2021}. However, behavioral diffusion and viral campaigns do not reliably generate social-tipping points, especially when the behavior is costly. 

A policy-first approach promotes a given behavior via a change in institutions such as laws, rules or other formal group-level procedures. These institutional changes may simply be rules promoting or requiring behavior, or they may entail behavioral support or enforcement. A policy approach can achieve rapid and large-scale behavioral change by altering the choices faced by an entire group. The scale and strength of the policy can determine the scale of adoption of the target behavior to a great extent. However, incentives can undermine intrinsic motivation for the behavior \cite{pellerano_2017, murayama_2010, promberger_2013}. Moreover, behavioral requirements can be perceived as coercive even when they result from collective democratic policy-making, as has been evidenced with the COVID-19 pandemic \cite{Schmelz_2021}. Consequently, policy based on behavioral requirements can fail if people stop participating when the enforcement or incentive is withdrawn. Thus, policy-first solutions are often powerful but brittle.

One critical aspect of social change that is often overlooked are the feedbacks between individual behavior and group-level policies \cite{Schmelz_Bowles_2021, Waring_CGS}. There are multiple ways in which policies, whether voluntary or coercive, can cause self-defeating social dynamics among citizens and `crowd out' beneficial behavior. Those reactions depend upon the culture of the group, which in turn has been conditioned by previous policies and institutions \cite{Schmelz_2021}. Therefore it is not possible to evaluate policy efficacy without understanding its reciprocal interactions with individual traits and behavior.

We think it is useful to consider behavioral transmission and institutional change as entangled parts of a larger process of social change. Social change is attractive as a policy framework because it combines the benefits of endogenous (bottom-up) behavioral change and (top-down) interventions and policy, and allows us to ask when and how policy and behavior come to be mutually reinforcing. However, unlike traditional policy frameworks, social change is at least partly an endogenous process, and cannot simply be implemented. In this paper, we ask how social change occurs when behaviors and institutions are mutually determined. We present a complex mathematical model for studying endogenous social change that couples simple mechanisms for the evolution of institutional policies and for the transmission of individual behaviors.

\section{Linking behaviors and institutions} 

Behaviors and institutions are similar phenomena on different social scales; both are cultural traits. Behavior, in most cases, is an individual cultural trait, or the expression of one, and institutions are group-level cultural traits. Both are invented, learned, modified, copied, and re-transmitted, diffusing and evolving within a larger population. Thus, the simplest method to unify behavioral change and institutional change is to consider their transmission in parallel \cite{Bowles_2003}.

\subsection{The transmission of behavior}

Behavioral diffusion has been studied from multiple perspectives, including the diffusion of innovation \cite{rogers2010diffusion}, behavioral contagion \cite{centola2010spread, centola2007complex}, and cultural transmission \cite{boyd_culture_1985}. The transmission of cultural traits via social learning, imitation and teaching is central to the process of cultural evolution, and helps drive linguistic change, economic development, and technological accumulation \cite{Beinhocker_2006, Boyd_2017, Henrich_2006}. Other cultural traits including the language, preferences, identities, beliefs and values that determine behavior are also transmitted via social learning between individuals \cite{Bandura_1971}. This social science is complemented by research in disease modeling on behaviors that respond to, and spread with, an infectious disease \cite{marceau2011modeling, hebert2020spread}.

Cooperative and pro-social behaviors are especially important, underlying successful group-level collective action by producing benefits beyond the actor. Behavioral science shows that humans are conditional cooperators \cite{Fischbacher_2001}, cooperating when either the situation \cite{Croson_Fatas_Neugebauer_2005} or the social institutions support cooperative behavior \cite{Baldassarri_2011}. Moreover, cooperative and pro-social behavior may sometimes be culturally transmitted between individuals within and between groups \cite{gachter_social_2005, molleman_cultural_2013, bergin_cooperation_2009}. However, because cooperative behaviors often come at a cost to the actor, they rarely persist or spread on their own. The factors that promote the evolution of cooperation have also been heavily studied. These mechanisms have two common features: they are payoff-modifying in that they make cooperation less costly or non-cooperation more costly, and group-focused in that they operate within some group of individuals somehow defined. Cooperation stabilizing factors include decentralized individual-level mechanisms to exclude non-cooperators (e.g. ostracism) or reduce their payoffs (e.g. peer punishment), and centralized group-level institutions (e.g. pooled-punishment, or policing) which also change payoffs. 

Since cooperation grows best in groups, inter-group processes such as competition and selection typically spur the evolution of cooperation in certain regards. Cooperation and institutional traits that support it can spread when human groups compete \cite{Bowles_2003, turchin_ultrasociety_2016, richerson_cultural_2016, van_den_bergh_group_2009}. For example, trust tends to grow to higher levels in more competitive industries \cite{francois_2018}. Game theoretic and evolutionary models show how cooperation can spread through group-level selection even when individuals face a social dilemma with other group members \cite{traulsen_evolution_2006, garcia_evolution_2011, simon_towards_2013}. Analysis of these models reveals that a certain critical threshold of group-level pressure is required to make cooperation evolve generally. This cultural group selection mechanism helps explain both the spread of cooperation and the emergence of institutions that support it \cite{Bowles_2003, garcia_evolution_2011, Waring_CGS, Gavrilets_2021}.

\subsection{The evolution of institutions}

Institutions, including centralized mechanisms for punishment, redistribution, division of labor and collective decision-making are a type of group-level cultural trait \cite{Smaldino_2014}. In comparison to decentralized behavioral regimes such as peer punishment or ostracism, institutions are widely considered the most effective and important mechanisms by which human groups stabilize cooperative behavior. Institutions can stabilize cooperation and drive wide-spread behavioral adoption because they can render a cooperative behavior individually beneficial by changing the outcomes and payoffs from each behavioral option. Institutional evolution is therefore believed to  gradually increase the functionality of institutional arrangements for the groups that adopt them. For example, the most studied institutional features, Ostrom's (1990) institutional rules \cite{ostrom_governing_1990}, are considered to be group-beneficial cultural adaptations which evolved in large part to help groups maintain cooperation in important domains of shared interest \cite{wilson_2013}. Thus it is believed that institutions emerge and spread largely because they enable groups to resolve collective action problem and social dilemmas \cite{Henrich_2006, richerson_cultural_2016}. High levels of institutionalization, such as are found in rich industrial societies, can support a cooperative and productive economic system.

Institutions also spread between groups, via processes largely parallel to the mechanisms of social learning at the individual level. Research on policy diffusion shows that the spread of policy, rules, and institutions between groups is also an important determinant of change \cite{Tews_2003, Shipan_Volden_2008}. Institutions may spread between companies, nations, municipalities, sports teams, or in any population of human groups. So, naturally, institutional evolution depends on the transmission of institutions between groups \cite{Hodgson_2002}. Thus, when organizations, nations, and companies may imitate, modify and re-transmit helpful institutions, the result is the evolution of organizational structure, business models, municipal rules and national law. In fact, the concurrent spread of behavior change and institutional support was recently identified as a needed avenue of research in public health modeling \cite{bedson2021review}. But currently, the transmission of individual-level behavior and group-level institutions are typically considered in isolation.

Therefore, the relationship between the diffusion of behavior and the spread of group policies and institutions is important for societal outcomes, yet remains understudied. On one hand cooperative behavior only tends to spread when institutions support it and render it less costly or altruistic for participants. On the other, institutions themselves only emerge when a group has a shared identity and intention to resolve a common goal: when cooperation is not a problem. So, the cooperation-institutions issue is a chicken-and-egg question; to understand one, we must understand both.

To help consider how policy and behavior can produce beneficial societal outcomes, we ask how behavior and institutions interact in a group-structured population. We develop a model of the coupled diffusion of individual behavior and group-level institutions using a novel approach that combines epidemiological (e.g. \cite {Bettencourt_2006}) and evolutionary approaches (e.g. \cite{Gavrilets_2021}). We are interested in cooperative or pro-social behavior which will dissipate without group-level support, but that can spread when a group adopts a policy to promote it. The greater the group's efforts to promote the behavior, the more rapidly it spreads within that group. We couple the diffusion of this behavior with a process of group-level institutional evolution, in which groups with more of the behavior achieve better group-level outcomes. Groups with better outcomes are preferentially imitated, leading to the spread of institutions. We explore how this coupled system of behavioral and institutional evolution operates as a whole. The model provides some useful insights on the nature of social change, and may help design new ways to achieve large societal goals more effectively.

\section{Model} 

\begin{figure*}
    \centering
    \includegraphics[width=\linewidth]{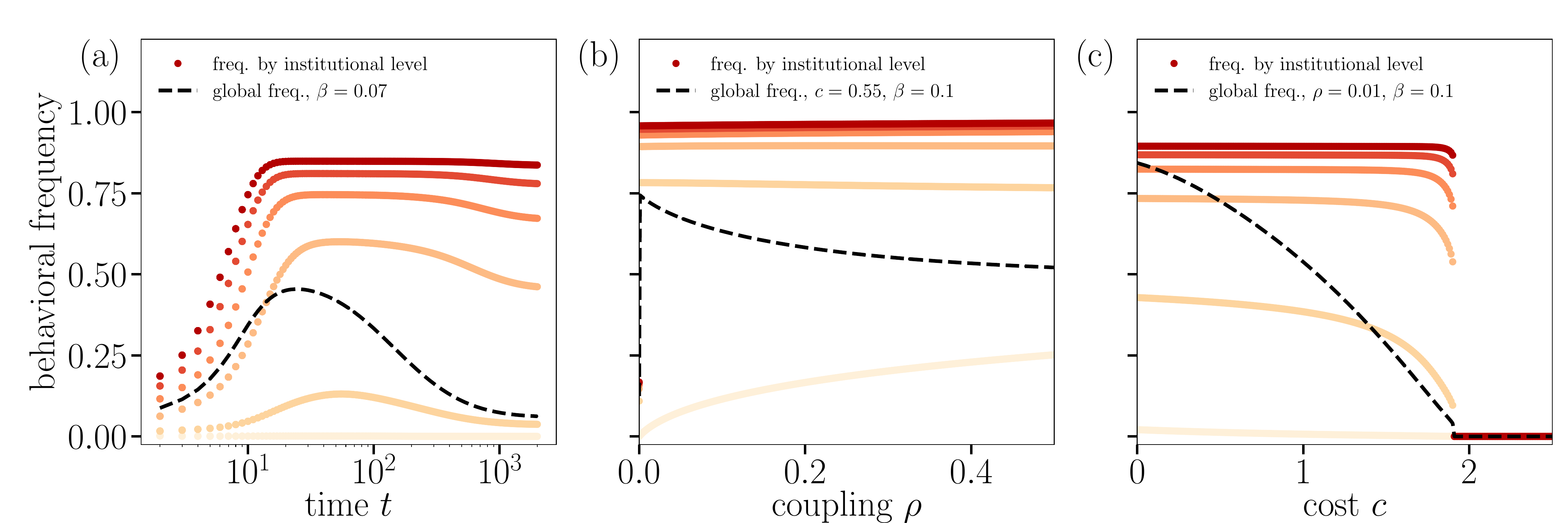}
    \caption{\textbf{Frequency of behavior in groups with different institutional strength.} 
      Within groups, the frequency of cooperative behavior follows the strength of institutions, shown in shades of red ($\ell =0$ in light beige and $\ell = 5$ in dark red). \textbf{(a)} The time dynamics of global behavioral frequency (black dashed line) and behavior in groups can include patterns of surge and collapse. \textbf{(b)} Increasing global diffusion of behavior, $\rho$, does not affect the frequency of behavior in institutionalized groups but increases adoption of the behavior in groups that have no institution (light beige). This shift increases the relative fitness of weaker institutions, leading to a decrease in global adoption of cooperation (see figure \ref{fig:institutions}). \textbf{(c)} Increasing the cost of institutions steadily decreases the global frequency of behavior, eventually reaching an inflection point (here $~1.9$) above which increasing costs causes synchronized discontinuous collapse of cooperation across all groups. Other parameters are fixed at $n=20$, $\gamma =1.0$, $b=0.18$ and $\mu = 10^{-4}$. }
    \label{fig:adoption}
\end{figure*}

We assume a population of size $N \gg 1$, divided into $M$ groups of equal size $n = N/M$. Groups contain $i$ adopters of the behavior and $(n-i)$ non-adopters. The natural diffusion rate, $\beta$, is the rate at which the behavior spreads from adopters to non-adopters within groups at a basic institutional level that allow but neither promote nor discourage the behavior. Adopters also lose that behavioral trait at a rate $\gamma$. Groups adopt an institution from a series of discrete levels of institutional strength, $\ell$, which modify the rate of behavioral diffusion to be $\ell\beta$. Note that without group-level institutions --- i.e., at $\ell=0$ --- the behavior does not spread at all. Therefore, like behaviors that benefit the group but incur individual costs such as cooperation and altruism, the behavior is likely to dissipate if not also adopted by one's social contacts. The institutional strength of a group can therefore be conceptualized as setting the susceptibility of individuals in that group to adopting the behavior. Finally, global behavioral diffusion, $\rho$, allows the behavior to spread between groups, based on the institutional strength of the receiving group. 

We then implement this system of behavioral diffusion within and among groups through a set of master equations \cite{hebert2010propagation} that tracks the fraction of groups $G_{i,\ell}(t)$ with $i$ adopters and institutional level $\ell$. Omitting the time dependency for simplicity, the behavior diffuses following

\begin{equation} \label{eq:diffusion_me}
    \begin{split}
    \frac{d}{dt}G_{i,\ell}^{\textrm{diff}} = & \phantom{+} \ell\beta\left[(i-1)+R\right]\left(n-i+1\right)G_{i-1,\ell} \\
    & - \ell\beta\left(i+R\right)\left(n-i\right)G_{i,\ell} \\ & + \gamma\left(i+1\right)G_{i+1,\ell}
     - \gamma i G_{i,\ell} \; . 
    \end{split}
\end{equation}

In this equation, the first term corresponds to diffusion events bringing groups from state $(i-1,\ell)$ to state $(i,\ell)$, which occurs proportionally to the internal diffusion rate $\ell\beta$ times the number of non-adopters $n-(i-1)$ who could adopt the behavior. The factor in square brackets corresponds to the number of all adopters with influence over the non-adopters within that group, which includes $i-1$ adopters within the group and the non-adopters in other groups who are exposed via global diffusion, i.e.
\begin{equation}
R = \rho \sum _{i',\ell'} i' G_{i',\ell'} \; .
\end{equation}
where primes simply denote variables over which we sum to calculate a global quantity.

\begin{figure*}
    \centering
    \includegraphics[width=\linewidth]{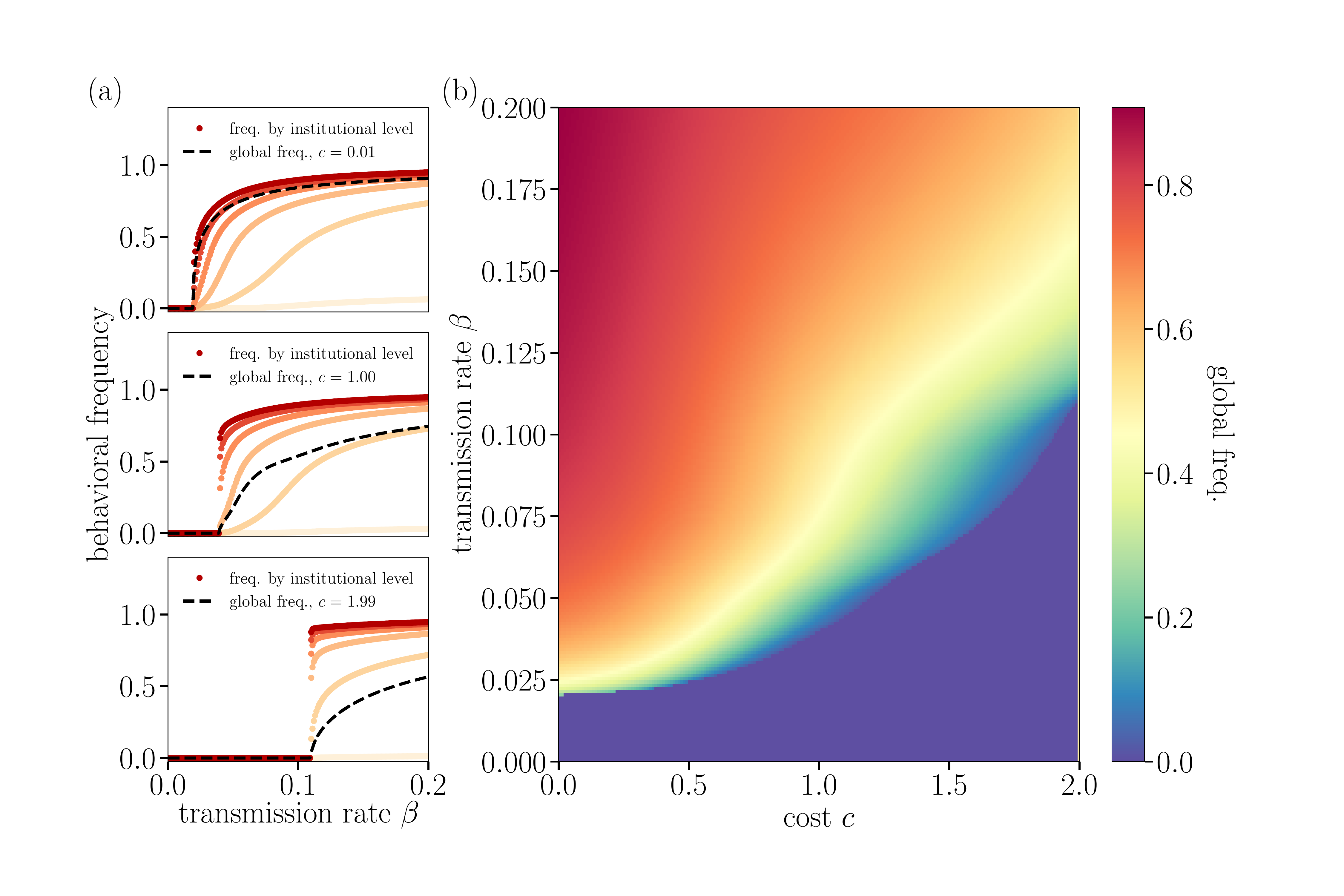}
    \caption{\textbf{Behavioral transmission and institutional cost drive behavior frequency.} Institutional costs ($c=0.01$, 1.0, 1.99) interact with behavioral transmission $\beta$ to determine frequency of cooperation behavior. \textbf{(a)} At lower costs behavior emerges gradually with increasing transmission, but at higher costs this occurs more rapidly and uniformly across groups at different institutional levels. In groups with institutions $\ell \geq 2$, behavior spreads extremely rapidly, while groups with $\ell = 1$ see a slower, and continuous activation with increasing $\beta$. Groups without institutions, $\ell = 0$, displayed as the lightest shade, can not generate cooperation but still see an increase in cooperation due to the global transmission from other groups. \textbf{(b)} This same interaction can be seen in the global frequency of behavior across all combinations of $\beta$ and $c$. Parameters fixed at $n=20$, $\gamma =1.0$, $b=0.18$ and $\mu = 10^{-4}$.}
    \label{fig:heatmap}
\end{figure*}

The second term in the master equation also corresponds to diffusion events but now taking groups out of state $(i,\ell)$ and into state $(i+1,\ell)$. Finally, the last two terms correspond to relaxation events where adopters revert to their old behavior; bringing groups from state $(i+1,\ell)$ to $(i,\ell)$ or from $(i,\ell)$ to $(i-1,\ell)$.

Additionally, we model institutional evolution among groups. Institutions are defined as a series of discrete levels of institutional strength $\ell \in [\ell_{\textrm{min}},\ell_{\textrm{max}}]$, here set as $\ell \in [0,5]$ but easily generalizable. Greater levels of institutional strength come with greater efficacy in promoting the group-beneficial behavior (i.e. $\ell\beta$) and with greater effort. Groups pay a cost $c$ for each additional level of institutional strength, and gain a collective benefit $b$ for each adopter in the group. Institutions evolve via a group selection process in which groups select their institutional strategy based upon the payoffs to groups with other strategies. We assume that a given level of institutional strength $\ell$ has a perceived fitness $Z_\ell$ given by

\begin{equation}
Z_\ell = \dfrac{\sum_{i'} \textrm{exp}\left(bi' - c\ell\right) G_{i',\ell}}{\sum_{i'} G_{i',\ell}}
\end{equation}

where the denominator acts as a normalization factor. Note that only the fitness of a group's strategy is directly observed and not the individual behavior of its members. Based on this perceived fitness, groups can change their institutional strength following a second master equation,

\begin{equation} 
    \label{eq:selection_me}
    \begin{split}
        \frac{d}{dt}G_{i,\ell}^{\textrm{select}} = & \rho \left[G_{i,\ell-1}\left(Z_\ell Z_{\ell-1}^{-1}+\mu\right) +G_{i,\ell+1}\left(Z_\ell Z_{\ell+1}^{-1}+\mu\right) \right] \\
        & -\rho \left(Z_{\ell-1}Z_{\ell}^{-1}+Z_{\ell+1}Z_{\ell}^{-1} + 2\mu\right)G_{i,\ell}\; . 
    \end{split}
\end{equation}

These terms track flows of groups from institutional strength $\ell$ to $\ell+1$ and $\ell-1$ (and vice versa). These flows are assumed to occur proportionally to the relative fitness of different levels (e.g. a group moves from $\ell$ to $\ell +1 $ proportionally to $Z_{\ell+1} / Z_{\ell}$) and the rate of global behavioral diffusion $\rho$, which is equivalent for individual behavior and group institutions. Importantly, we also add a constant rate of transition $\mu$ regardless of fitness which allows the invention of new institutional levels (e.g., discovery of unoccupied institutional strategies). 

The total dynamics of our model is then given by a set of master equations,

\begin{equation}
\frac{d}{dt}G_{i,\ell}= \frac{d}{dt}G_{i,\ell}^{\textrm{diff}} + \frac{d}{dt}G_{i,\ell}^{\textrm{select}} \; ,
\end{equation}

defined over $i \in [0,n]$ and a set of integers for levels of institutional strength $\ell \in [\ell_{\textrm{min}},\ell_{\textrm{max}}]$. We can thus track the dynamics of our system by numerically integrating all equations starting from arbitrary initial conditions (constrained only by $\sum_{i,\ell} G_{i,\ell} = 1$).

The system is sufficiently complex to rule out analytical descriptions of equilibria, but we explore stability conditions analytically in Appendix. We simulate systems with identical initial conditions, specifically a uniform distribution of groups across institutional levels and a binomial distribution of $i$ with average of 1\% adopters. Stable long-run values were recorded when $t \geq 10,000$ and if the expected average difference $\Delta i$ in groups was less than $10^{-10}$ over a $\Delta t = 1$.

\begin{figure*}
    \centering
    \includegraphics[width=\linewidth]{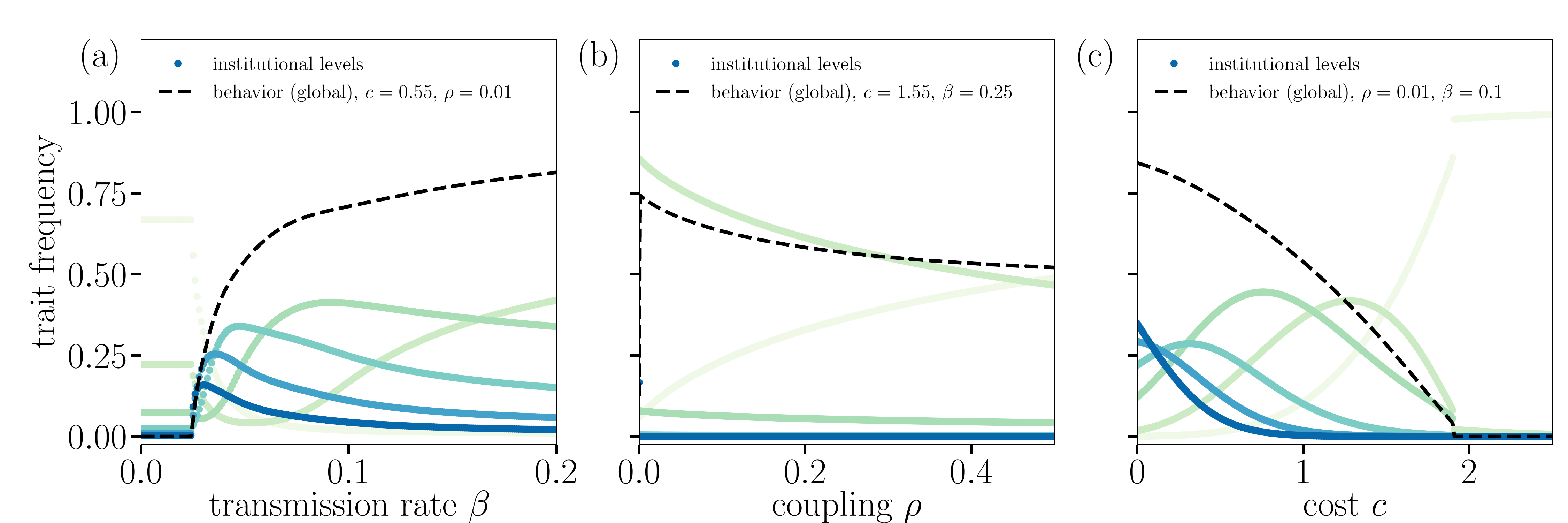}
    \caption{\textbf{Institutional frequency among groups.} The adoption of different levels of institution (i.e., $\sum_{i'}G_{i',\ell}$) is shown with $\ell =0$ in light green and $\ell = 5$ in dark blue). \textbf{(a)} At low values of $\beta$ groups with $\ell = 0$ are most common. As the transmission rate crosses the threshold, stronger institutions ($\ell>2$) proliferate rapidly, while weaker institutions first decline in frequency, then grow to become dominant. Institutions of $\ell=2$ or $\ell=3$ become most popular at intermediate values of $\beta$. Stronger institutions are progressively favored at lower value of cost $c$ and global diffusion $\rho$. \textbf{(b)} Increasing the global diffusion rate $\rho$ confirms the emergent institutional free riding problem: the no-institutions strategy progressively gains in popularity as global diffusion of cooperation increases, eventually becoming dominant. \textbf{(c)} Stronger institutions are favored at lower cost. As costs institutional increase lower levels of institution are progressively preferred. This dynamic continues until the lowest non-zero institutional level $\ell=1$ also loses its fitness at the cost threshold, where these groups experiences a discontinuous transition. Parameters fixed at $n=20$, $\gamma =1.0$, $b=0.18$ and $\mu = 10^{-4}$.}
    \label{fig:institutions}
\end{figure*}

\section{Results} 

\begin{figure*}
\centering
    \includegraphics[width=\linewidth]{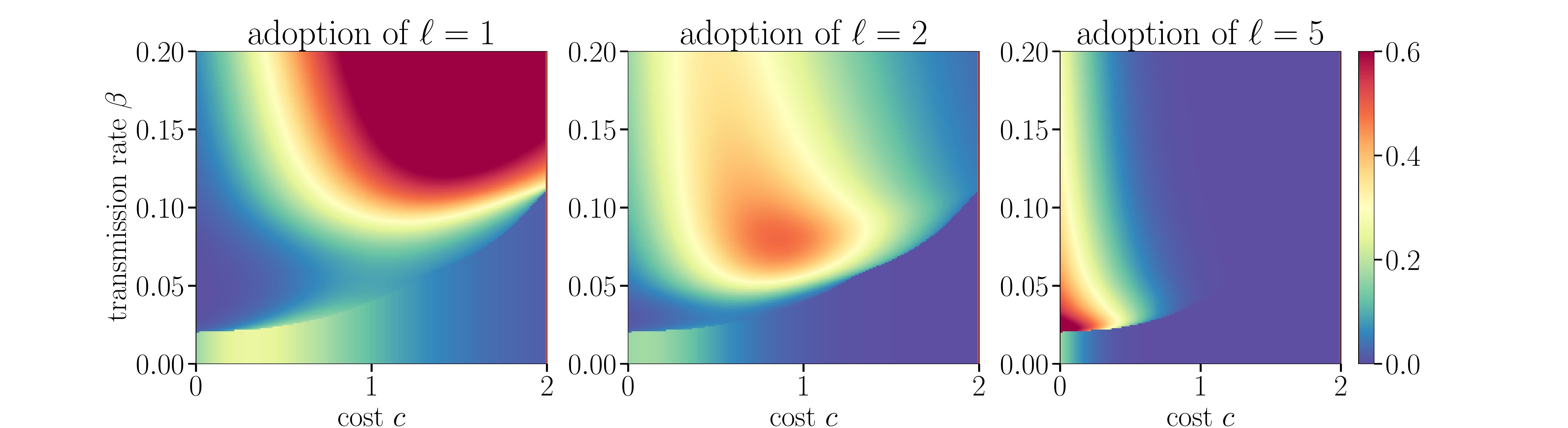}
    \caption{\textbf{The prevalence of institutions of different strengths varies by costs and transmission rate.} Panel replicates the numerical experiment of Fig.~\ref{fig:heatmap}, but shows the adoption of institutions of level 1, 2, and 5, respectively. This comparison highlights the phenomenon of institutional localization in which a given institutional level dominates the fitness landscape in some subset of parameter space. This phenomenon is analytically explored in the Appendix.}
    \label{fig:localization}
\end{figure*}

The model produces a set of co-evolutionary dynamics between behaviors and institutions, some of which were unexpected, and hold value as tools for understanding social change. We summarize model behavior in Figs.~\ref{fig:adoption}, \ref{fig:heatmap}, \ref{fig:institutions}, and \ref{fig:localization}. 

As expected, the model confirms that cultural group selection on institutions can drive evolution of cooperation or pro-social behavior, as others have demonstrated \cite{traulsen_evolution_2006, garcia_evolution_2011, Waring_CGS}. We find that for behavioral-institutional co-evolution to take off, two fundamental criteria must be met: institutional implementation costs, $c$, must be below a threshold value (see Fig.~\ref{fig:adoption}, and the rate of behavioral transmission within groups, $\beta$, must be above a critical threshold (see Fig.~\ref{fig:heatmap}.  These thresholds are sharp.  Even small changes in $c$ and $\beta$ can have catastrophic effects (see Fig.~\ref{fig:heatmap}). In fact, we find a rich mix of both continuous and discontinuous emergence of the behavior depending on institutional levels and parameter values. We explore the mathematical conditions around these transitions in the Appendix.

Qualitatively, no institutions are possible if institutional costs are too high, and the behavior never spreads. Lowering institutional costs slightly below the threshold results in a sharp discontinuity such that institutions and behavior both flourish. Further decreases do not change qualitative outcomes. A similar threshold also exists for within-group behavioral transmission, $\beta$. Unlike the cost threshold, however, increases in $\beta$ above the threshold result in further spread of both behavior and institutions. However, if costs are low, institutions may still evolve even with a very low $\beta$ (see Fig.~\ref{fig:institutions}a). 

Naturally, simple institutions (i.e. $\ell = 1$) emerge and become most common across a large range of parameters ($c$, $\beta$, $\rho$) (Fig.~\ref{fig:institutions}c). Stronger institutions (i.e. $\ell \geq 2$) also evolve if costs are sufficiently low enough (Fig.~\ref{fig:institutions}). However, as costs rise toward the threshold, the greatest level of institution that can be sustained declines steadily.  Note that our model is harsh on the evolution of stronger institutions, because we do not allow institutions to modify costs, which they are typically assumed to do. Therefore, we suspect that the model under-estimates institutional evolution. 

Under certain conditions the model exhibits an unexpected pattern of bi-modal institutional success. This occurs when a certain level of institutionalization is required to maintain the behavior locally, but only some groups become self-sustaining. For example, close to the sustainability thresholds, it is possible to arrive at a state in which $\sim20\%$ of groups have no institution, $\sim 25\%$ with level 3 institutions, but very few have level 1 or 2 institutions (Fig.~\ref{fig:institutions}a). This state emerges when the intervening institutional level ($\ell = 1$ or 2) has lower fitness than either $\ell = 0$ or $\ell = 3$, hindering non-institutional groups from moving to stronger institutions. We call this phenomenon institutional localization. We explore it visually in Fig.~\ref{fig:localization} and analytically in details in the Appendix.

The rate of global behavioral diffusion, $\rho$, also creates a counter-intuitive effect. As $\rho$ increases, total behavioral adoption and institutions both \textit{decrease} (see Fig.~\ref{fig:institutions} b). Increased global diffusion allows the cooperation to spread into groups that do not have institutions to generate or support it. Cooperation in receiving groups does not persist, but still increases their fitness. Higher fitness outcomes for poorly institutionalized groups then creates selection against stronger institutions. In essence, high between-group transmission, $\rho$, creates behavioral source-sink dynamics in which cooperation generated in one group spreads via imitation and learning to other groups improving the outcomes of those groups without requiring effective institutions. Put in evolutionary terms, high between-group transmission of cooperation can halt cultural group selection though the emergence of institutional free-riding.

The source-sink dynamics of cooperation have an important influence on model outcomes. In the region around the emergence threshold model behavior is complex and outcomes are sensitive to initial conditions. In this region, we find non-monotonic time series in which behavior and institutions can spread together for a time, only to collapse later (Fig.~\ref{fig:adoption}a). In certain parameter combinations, the model exhibits multiple stable equilibria, ending either in the collapse of behavior and institutions (e.g. all groups in $i=\ell=0$) or with a steady state non-zero level of both behavior and institutions, depending finely on initial conditions. This model thus provide an interesting mathematical sandbox for future work considering how to design promotion of beneficial collective behavior and institutional support.  


\section{Conclusion}
Our model helps us understand social change when behavior and institutions are mutually linked. Connecting the diffusion of group-beneficial behavior to the spread of policy and supporting institutions reveals a set of social dynamics that are not often studied. Our analysis shows that group-beneficial behavior can be subject to source-sink dynamics in which cooperation generated by groups with strong institutions spreads to less-institutionalized groups. These behavioral source-sink dynamics create an emergent pattern of institutional free-riding in which receiving groups benefit from the imported cooperative behavior, but cooperation dies away in those groups without institutional support. Thus, receiving groups act as a cooperation sink. We also find that when the global spread of cooperation, $\rho$, is sufficiently strong, these dynamics can halt the evolution of institutions. These dynamics can dampen or halt cultural group selection for institutions when the behavioral diffusion increases the fitness of receiving groups with weak institutions.

Source-sink dynamics of cooperation may be influencing social evolution all the time, and examples abound. One area of interest is when the provision of public goods is difficult to understand or perceive. When the cooperation of one group creates benefits for other groups, institutional free-riding may result. One example of this may occur in national tax systems when tax revenue is shared between regions. If such revenue sharing enhances the perceived success of recipient states, it may also undermine institutional evolution among states.  Another immediate example concerns the heterogeneous and slow spread of robust public health policy for fighting COVID-19 in the United States \cite{Renne_2020, white2020state, althouse2020unintended}. States with robust policy (i.e. masking, social distancing and vaccination requirements) likely spread those beneficial behaviors to states with weak public health policy. This behavioral spillover could artificially increase the actual or perceived success of states with weak policies, slowing the uptake of effective policies around the country. 

Source-sink dynamics of cooperation might also help explain long-term patterns of social change, such as the decline of unions in the United States \cite{Hogler_2015}. Collective bargaining often improves work conditions and pay for union members, but can also benefit non-union members. For example, in right-to-work states, people can reap the rewards of a union without joining \cite{Sobel_1995}. This undercuts the strength of the local union, but it may also reduce union fitness generally, slowing their spread elsewhere. Thus, limiting negotiated benefits to members might improve the correlation between institution and behavior, helping to maintain both. Another example comes from food safety regulations. A lack of awareness of the role of the safety regulations can lead municipalities to repeal them. But, when people mostly consume food from safely regulated production systems, those non-regulated municipalities are protected from negative consequences, and the reduced safety standards can spread. Something similar has been observed in Maine, where reduced food safety standards have spread between many small towns \cite{Hupper_2019}. These examples suggest that policies that create public goods can be made more durable if they also make their supply of those public goods highly observable. We believe the source-sink dynamics of cooperation provide a new and useful framework and worthy of additional study. 

In combining behavioral transmission with institutional evolution the model shows that it is possible to overcome limitations of the policy-first and behavior-first approaches. Specifically, our model has implications for intervention efforts to achieve large societal goals that require social change which do not emerge from the standard approaches. First, we find that, despite reciprocal dynamics between behavior and policy, the spread of institutions has a greater influence on behavioral diffusion than the reverse. This is an important insight for both social scientists and policymakers in the age of nudges and viral behavioral campaigns: as a group-level cultural trait, policy has an out-sized influence on the evolution of a group-structured society. But social change is not owned and cannot be controlled by any one actor, even including legitimate governments. Any actor within society might look to influence social change. Indeed, beyond political domains many actors already intentionally and strategically influence social change. For example, for-profit corporations spend large sums of money in advertisements to modify consumer preferences and purchasing behavior. Therefore, it is useful to provide intervention points for actors in different social positions. We focus on two levels, first the level of the group (such as a state or organization) and second, the level of the population of groups (such as a nation).  

Group-level actors, such as state governments or organizations, have two major avenues for action that emerge from our framework. First, groups can work to improve the spread of the cooperative or beneficial behavior within the group. For example, create meaningful rewards for the group-beneficial behavior, or ways of celebrating the contributions of those who contribute. Second, groups can adopt new institutions (which enhance behavioral transmission) by learning from other innovative groups. Innovative institutions may be social in nature, such as finding and enhancing ways for individuals to reward, recognize and congratulate each other on their contributions. But groups should also be careful to avoid relying on externally generated cooperation, lest it cease to flow. This framework could useful for economic development in rural states.

Population-level actors, such as national governments, can work to improve the spread of policies that support beneficial behavior between groups. There are two main fronts in this effort. First, reducing the costs of implementing policies that support group-beneficial behavior can be very effective at spreading both behavior and policy. For example, a central government could provide financial support to cities that adopt a policy to support climate-friendly behavior. Second, increasing between-group learning of policy options can accelerate behavior-policy co-evolution. For example, the U.S. Federal Government could enhance the transmission of beneficial state policies by supporting conferences between state governments to facilitate the exchange of ideas and solutions in a domain of interest (e.g. combating addiction).

Actors at the population level must also contend with source-sink dynamics of behavior which can undermine cooperation and institutional evolution. This process is determined in part by the rate of transmission of cooperation between groups. Therefore, our model suggests that efforts to spread cooperation globally are likely to be somewhat self defeating. Furthermore, we would caution against strategies for reducing the transmission of cooperation between groups, as such policies could counteract the transmission of cooperation \textit{within} groups, which is of even greater importance. Instead, population level actors should work to create a pattern of reinforcement between cooperation behavior and local institutions. One way to accomplish this is through correlated pilot projects: efforts to boost the \textit{correlated} spread of beneficial behavior and supporting institutions. In our model, behavior and institutions spread best when they are correlated across groups and with beneficial outcomes. Therefore, policy approaches might focus on promoting strong pilot studies and local public trials with high chances of success to be copied elsewhere. Future research could use this model to study examples of pilot project diffusion. For example, a study of coastal management pilot projects in South Africa suggests that the coupled diffusion of cooperation and institutions may be useful for achieving societal transitions to more sustainable regimes \cite{Vreugdenhil_2012}. It seems likely that similar dynamics might underlie the diffusion of environmental regulation \cite{Tews_2003, Busch_2005, Daley_2005}.

Our model sheds light on a set of social processes that may be very influential in social evolution. If the dynamics that we demonstrate here are as prevalent as they appear, the implications of this study are deep and far-reaching. Moreover, our model provides important insights on the efficacy of approaches for addressing global challenges such as climate change. It provides a novel framework to study intervention regardless of the social scales at which actors operate.


\section*{Acknowledgements}
This research was supported by the National Science Foundation [award EPS-2019470], Google Open Source under the Open-Source Complex Ecosystems And Networks (OCEAN) project, and the Department of Agriculture [NIFA project \#ME022008]. The authors thank Adrian Bell, Jeremy Brooks and Vicken Hillis for inspiration. 

\appendix

\section{Group fitness distribution}

\begin{figure*}
    \centering
    \includegraphics[width=\linewidth]{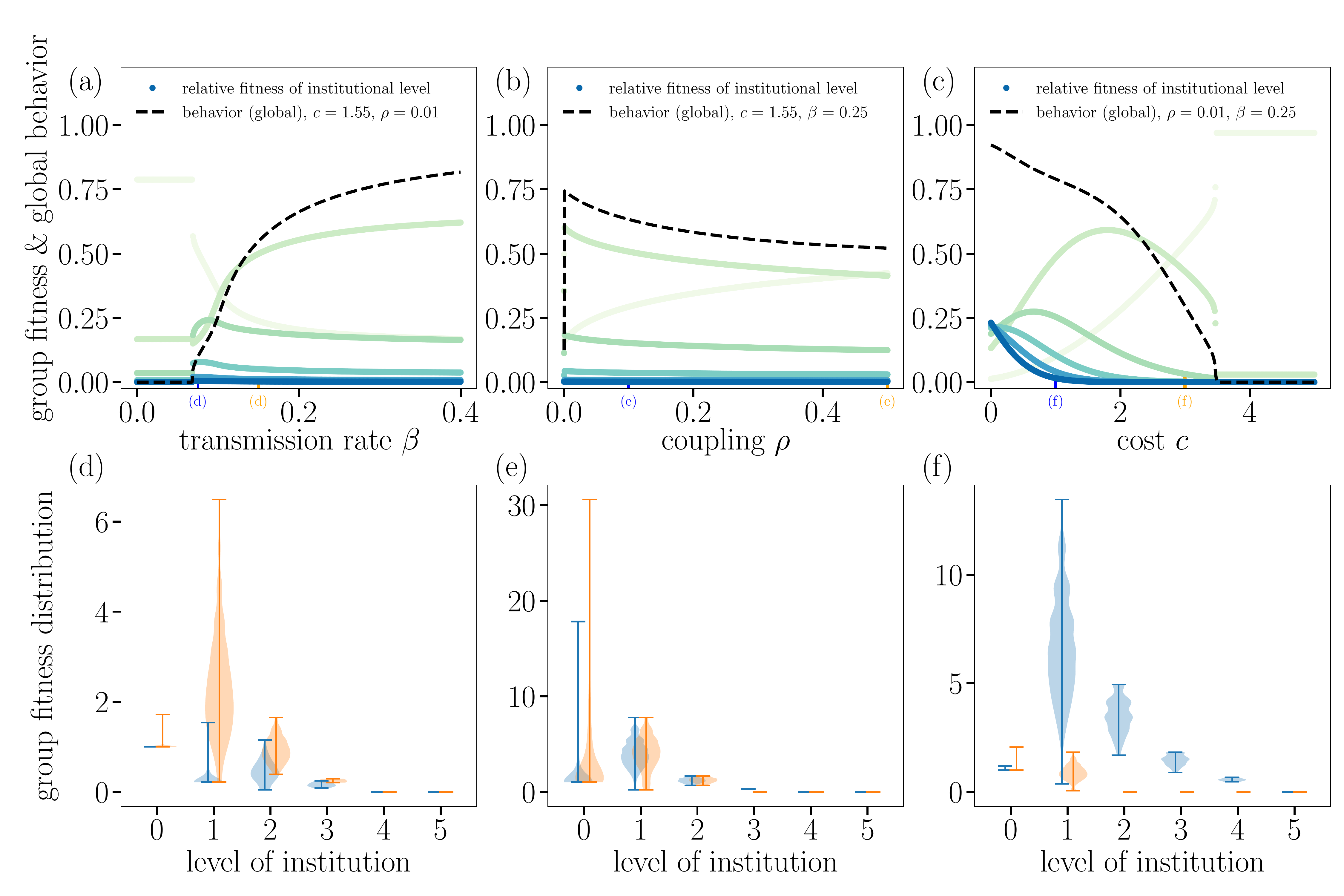}
    \caption{\textbf{Group fitness across institutional levels.} Top row \textbf{(a-c)} replicates results of Fig.~\ref{fig:institutions} but shows the average fitness of different institutional levels rather than their frequency. Markers on the horizontal axis show parameters used in the second row. The second row, \textbf{(d-f)} shows full fitness distribution (i.e., not averaged over the local number $i$ of adopters of the behavior) between parameter values. Altogether, this illustrates that average fitness is a relatively good description of group traits and trait frequencies.}
    \label{fig:fitness}
\end{figure*}

In Fig.~\ref{fig:fitness}(top row) we reproduce some of the results of Fig.~\ref{fig:institutions} but instead looking at the average group fitness for groups with a given level of institutions, rather than looking directly at the adoption (or trait frequency) of that institutional level. This confirms that, as expected, fitness and frequency of group institutional levels are proportional and essentially equivalent. More interestingly, Fig.~\ref{fig:fitness}(bottom row) looks at the distribution of fitness across all local behavioral frequencies for groups at a given institutional level. These results help explain why trait frequency and average fitness are directly proportional: The underlying distributions of fitness can be broad but are always unimodal and well described by their average.

\section{Localization of adoption across levels of institutional strength}


For certain parameters, the adoption of cooperative behavior is very heterogeneous across levels $\ell$.
This can be explained by the emergence of \textit{localization} in groups with a certain institutional level \cite{stonge2021social}.
We characterize the phenomenon using
\begin{align}
    A_\ell = \frac{\sum_i \frac{i}{n} G_{i,\ell}}{\sum_i G_{i,\ell}} \;,
\end{align}
the adoption of innovation in groups of institutional level $\ell$.
In a nutshell, if the adoption is \textit{localized} in level $\ell$ but not $\ell'$, then we have $A_\ell \gg A_{\ell'}$.

To illustrate how this phenomenon arises, we make use of a perturbative approach in the stationary state ($t \to \infty$) to characterize $A_\ell^*$.
To simplify the steps, we introduce a proportionality constant $\epsilon$ in front of the institutional level flow different from the global diffusion of behavior, $\rho$.
Our perturbative development is valid when $\epsilon \ll 1$ and $R \to 0$, so that the selection process happens at a much slower rate than the diffusion and we are near the absorbing-state (no adoption in the population).

First, let us rewrite Eqs.~\eqref{eq:diffusion_me} and \eqref{eq:selection_me} using marginal $G_{\ell} = \sum_i G_{i,\ell}$ and conditional $G_{i|\ell} = G_{i,\ell}/G_\ell$ probabilities, normalized as $\sum_i G_{i|\ell} = 1$ and $\sum_\ell G_\ell = 1$.
We can rewrite the fitness as
\begin{align}
    Z_\ell = \frac{\sum_i e^{b i - c\ell} G_{i|\ell}}{\sum_i G_{i|\ell}} \;.
\end{align}
and
\begin{align}
    A_\ell = \frac{1}{n}\sum_i i G_{i|\ell}
\end{align}

On the one hand, the marginal probability $G_{\ell}$ evolves according to
\begin{align}
    \frac{\mathrm{d}}{\mathrm{d}t} G_\ell &= \sum_{i} \frac{\mathrm{d}}{\mathrm{d}t} G_{i,\ell}^{\mathrm{diff}} + \frac{\mathrm{d}}{\mathrm{d}t} G_{i,\ell}^{\mathrm{select}} \;, \notag \\
                                       &= \epsilon \left [  G_{\ell-1} \left ( Z_\ell Z_{\ell-1}^{-1} + \mu \right) + G_{\ell+1} \left ( Z_\ell Z_{\ell+1}^{-1} + \mu \right) \right] \notag \\ &\phantom{=} - \epsilon G_\ell \left ( Z_{\ell+1} Z_{\ell}^{-1} + Z_{\ell-1} Z_{\ell}^{-1} + 2\mu \right) \;,
\end{align}
where at the second line we make use of the fact that the diffusion terms $\frac{\mathrm{d}}{\mathrm{d}t} G_{i,\ell}^{\mathrm{diff}}$ cancel each other.
In other words, the marginal probability $G_\ell$ is mainly driven by the \textit{selection} process.

In the stationary state ($t \to \infty$), the stationary probability $G_\ell^*$ must respect \textit{detailed balance},
\begin{align}
    G_{\ell+1}^*(Z_\ell^* /Z_{\ell+1}^* + \mu) = G_\ell^* (Z_{\ell+1}^*/ Z_\ell^* + \mu) \;,
\end{align}
i.e., the probability flow between states $\ell$ and $\ell+1$ must balance each other.
Therefore, we have as a general solution
\begin{align}
    \label{eq:Gl_gen_sol}
    G_\ell^* = G_0^* \prod_{\ell' = 0}^{\ell-1} \frac{Z_{\ell+1}^*/ Z_\ell^* + \mu}{Z_\ell^*/ Z_{\ell+1}^* + \mu} \;.
\end{align}
On the other hand, the conditional probability $G_{i|\ell}$ follows
\begin{align}
    \frac{\mathrm{d}}{\mathrm{d}t} G_{i|\ell} = \frac{1}{G_\ell}\frac{\mathrm{d}}{\mathrm{d}t} G_{i,\ell} - \frac{G_{i|\ell}}{G_\ell} \frac{\mathrm{d}}{\mathrm{d}t} G_\ell \;.
\end{align}
While it is possible to rewrite the equation above in terms of $G_{i|\ell}$ and $G_\ell$ and solve numerically for the stationary states $G_{i|\ell}^*$ and $G_\ell^*$, we want to extract some analytical insights from the master equations.
We therefore consider the following perturbative developments
\begin{align}
    G_{i|\ell} = G_{i|\ell}^{(0)} + G_{i|\ell}^{(1)} \epsilon + O(\epsilon^2) \;, 
\end{align}
The order zero term must still respect the normalization conditions $\sum_i G_{i|\ell}^{(0)} = 1$.
From now on, we consider that all quantities have reached the stationary state and we drop the asterisks to simplify the notation.
To order zero, the conditional probability must respect
\begin{align}
    0 = & \ell\beta\left[(i-1)+R\right]\left(n-i+1\right)G_{i-1|\ell}^{(0)} - \gamma i G_{i|\ell}^{(0)} \notag \\ &  - \ell\beta\left(i+R\right)\left(n-i\right)G_{i|\ell}^{(0)} + \gamma\left(i+1\right)G_{i+1|\ell}^{(0)} \;,
\end{align}
with solution
\begin{align}
    & G_{i| \ell}^{(0)} = G_{0| \ell}^{(0)} \binom{n}{i} (\lambda \ell)^i \prod_{j = 0}^{i-1} \left[j + R \right]  \quad \forall \;i > 0 ; \notag \\ & G_{0| \ell}^{(0)} = 1 - \sum_{i > 0} G_{i| \ell}^{(0)} \;,\label{eq:Gil_stat}
\end{align}
where $\lambda \equiv \beta/\gamma$.
Contrarily to the marginal probability, we see that the conditional probability $G_{i|l}$ is (up to order zero in $\epsilon$) mainly driven by the diffusion process.


Still, $G_{i| \ell}^{(0)}$ needs to be solved numerically since $R$ depends on $G_{i| \ell}^{(0)}$ and $G_l$.
We therefore consider the limit $R \to 0$.
Note that this limit is only possible if a stationary solution exists such that $R \to 0$, which is always the case for continuous phase transition and possible for certain discontinuous phase transition.
This allows us to further develop the solution as
\begin{align}
    G_{i| \ell}^{(0)} = \delta_{i,0} + g_{i|\ell}^{(0)} R + O(R^2)\;,
\end{align}
where $\delta_{i,j}$ is the Kronecker delta. Indeed, in the limit $R \to 0$, we have no adoption in the population, hence we must have $G_{i| \ell}^{(0)} \to \delta_{i,0}$ for all $\ell$.
Taking the derivative with respect to $R$ at Eq.~\eqref{eq:Gil_stat} around $R = 0$, we obtain for $\ell > 0$
\begin{align}
    g_{i|\ell}^{(0)} & = \binom{n}{i} \Gamma(i) (\lambda l)^{i} \quad \forall \; i > 0 ; \notag \\ g_{0|\ell}^{(0)} & = -\sum_{i > 0} g_{i|\ell}^{(0)} \;,
\end{align}
and for $\ell = 0$
\begin{align}
    g_{i|0}^{(0)} = 0 \quad \forall \; i \geq 0 \;.
\end{align}

We can now estimate the adoption for each institutional level $\ell$, $A_\ell$, using the perturbative development
\begin{align}
    A_\ell &= A_\ell^{(0)} + O(\epsilon) \;, \notag\\
           &= \frac{1}{n} \sum_i i G_{i|\ell}^{(0)} + O(\epsilon) \;, \notag \\
           &= \frac{1}{n} \sum_i i g_{i|\ell}^{(0)} R + O(\epsilon R + R^2) \;, \notag \\
           &= a_\ell^{(0)} R + O(\epsilon R + R^2) \;,
\end{align}
where we defined
\begin{align}
    a_\ell^{(0)} = \frac{1}{n} \sum_i i g_{i|\ell}^{(0)} \;.
\end{align}
Hence, for small $\epsilon$ and $R$, we can characterize $A_\ell$ through $a_\ell^{(0)}$ and thus through $g_{i|\ell}^{(0)}$.

For $\ell = 0$, we directly have $a_\ell^{(0)} = 0$, but for $\ell > 0$, the behavior is slightly more complicated.
Let us define the following generating function 
\begin{align}
    H_\ell^{(0)}(x) &= \sum_{i = 0}^n g_{i|\ell}^{(0)} x^i \;, \\
                    &= g_{0|\ell}^{(0)} + \sum_{i = 1}^n \binom{n}{i} \Gamma(i) (\lambda \ell x )^i \;, \notag \\
                    &= g_{0|\ell}^{(0)} + \int_0^\infty \sum_{i = 1}^n \binom{n}{i} (\lambda \ell x u)^i u^{-1} e^{-u} \mathrm{d} u \;, \notag \\
                    &= g_{0|\ell}^{(0)} + \int_0^\infty [(1 + \lambda \ell x u)^n-1] u^{-1} e^{-u} \mathrm{d} u \;, \notag
\end{align}
where at the third line we used the definition of the gamma function and at the fourth line we used the binomial theorem.
We can write $a_\ell^{(0)} = n^{-1}\partial_x H_\ell^{(0)}(x)|_{x = 1}$, thus
\begin{align}
    a_\ell^{(0)} &= \lambda \ell \int_0^\infty (1 + \lambda \ell u)^{n-1} e^{-u} \mathrm{d} u \;. \notag \\
                 &= \lambda \ell n \int_0^\infty e^{n \phi(y)} \Phi(y) \mathrm{d} y \;,
\end{align}
where $\phi(y) \equiv \ln(1 + \tilde{\lambda} \ell y) - y$, $\Phi(y) \equiv (1 + \tilde{\lambda} \ell y)^{-1}$, and $\tilde{\lambda} \equiv \lambda n$.
For sufficiently large $n$, we can approximate $a_\ell^{(0)}$ using standard asymptotic expansion of integrals \cite{olver1997asymptotics}.
If there is a critical point $u'$ on the interval $[0,\infty)$, we use Laplace method, otherwise we can simply develop $\phi(y)$ to first order and $\Phi(y)$ to order zero near $y = 0$.
We obtain
\begin{align}
    \label{eq:asymptotic_al}
    a_\ell^{(0)} \sim \begin{dcases}
        \frac{\lambda \ell }{1 - \tilde{\lambda} \ell} & \text{if } \tilde{\lambda}  <  \ell^{-1} \\
        \lambda \ell \sqrt{2 \pi n} e^{\psi n} & \text{if } \tilde{\lambda}  > \ell^{-1}
                      \end{dcases} \;,
\end{align}
where $\psi = \ln \tilde{\lambda} \ell - 1 + (\tilde{\lambda} \ell)^{-1}$, which means $\psi > 0$ if $\tilde{\lambda}  > \ell^{-1}$.

Equation \eqref{eq:asymptotic_al} gives us insight on when we should expect localization near the invasion threshold $\beta_\mathrm{c}$: If we have $\tilde{\lambda}_\mathrm{c} \equiv \beta_\mathrm{c} n /\gamma  \ll  L^{-1}$, then we can consider only the first case in Eq.~\eqref{eq:asymptotic_al}, and $a_\ell^{(0)} \propto \ell$ for all $\ell$---the adoption in all levels is proportional to $\ell$, but it is not dominated by any of them.
We call this regime \textit{delocalized}.
But if instead we have $\tilde{\lambda}_\mathrm{c}  >  \ell^{-1}$, then $a_\ell^{(0)} \propto \ell \sqrt{n} e^{\psi n}$, which is much larger then $a_\ell^{(0)}\propto \ell$. The adoption of innovation is thus concentrated in the institutional levels $\ell$ such that $\tilde{\lambda}_\mathrm{c}  >  \ell^{-1}$. 
In this regime, we say that the system is localized, and the levels $\ell \in \mathcal{L} \subseteq \lbrace 1,\dots,L\rbrace$ respecting the constraint $\tilde{\lambda}  >  \ell^{-1}$ near the invasion threshold is called the \textit{localization set}.


To investigate when a localization regime is possible, we must look at the invasion threshold of the system, which corresponds to the point where $R \to 0$---this is the critical point separating the absorbing phase from the active phase.
Recall that
\begin{align}
    R = \rho \sum_{i,\ell} i G_{i,\ell} = \rho \sum_{i,\ell} i G_{i|\ell} G_\ell \;.
\end{align}
Using our perturbative development, we write
\begin{align}
    R &= \rho \sum_{i,\ell} i \left [ G_{i|\ell}^{(0)} + G_{i|\ell}^{(1)}\epsilon \right]  G_\ell \;, \notag \\
      &= \rho \sum_{i,\ell} i  G_{i|\ell}^{(0)}  G_\ell + O(\rho \epsilon) \;,
\end{align}
Moreover, in the limit $R \to 0$ we obain
\begin{align}
    R &= \rho \sum_{i,\ell} i (\delta_{i,0} + g_{i|\ell}^{(0)} R)  [G_\ell]_{R = 0} + O(\rho \epsilon R + \rho R^2) \;, \notag
\end{align}
therefore, in the limit $R \to 0$, we have the constraint
\begin{align}
    1 = \rho \sum_{i,\ell} i g_{i|\ell}^{(0)} [G_\ell]_{R = 0} + O(\rho \epsilon) \;.
\end{align}
To order zero in our perturbative development, and since the sum on $i g_{i|\ell}^{(0)}$ corresponds to $n a_\ell^{(0)}$, we have
\begin{align}
    1 = \rho n \sum_{\ell} a_\ell^{(0)} [G_\ell]_{R = 0}  \;,
\end{align}
hence we only need to evaluate $[G_\ell]_{R = 0}$.

\begin{figure}[ht!]
\centering
    \includegraphics[width=\linewidth]{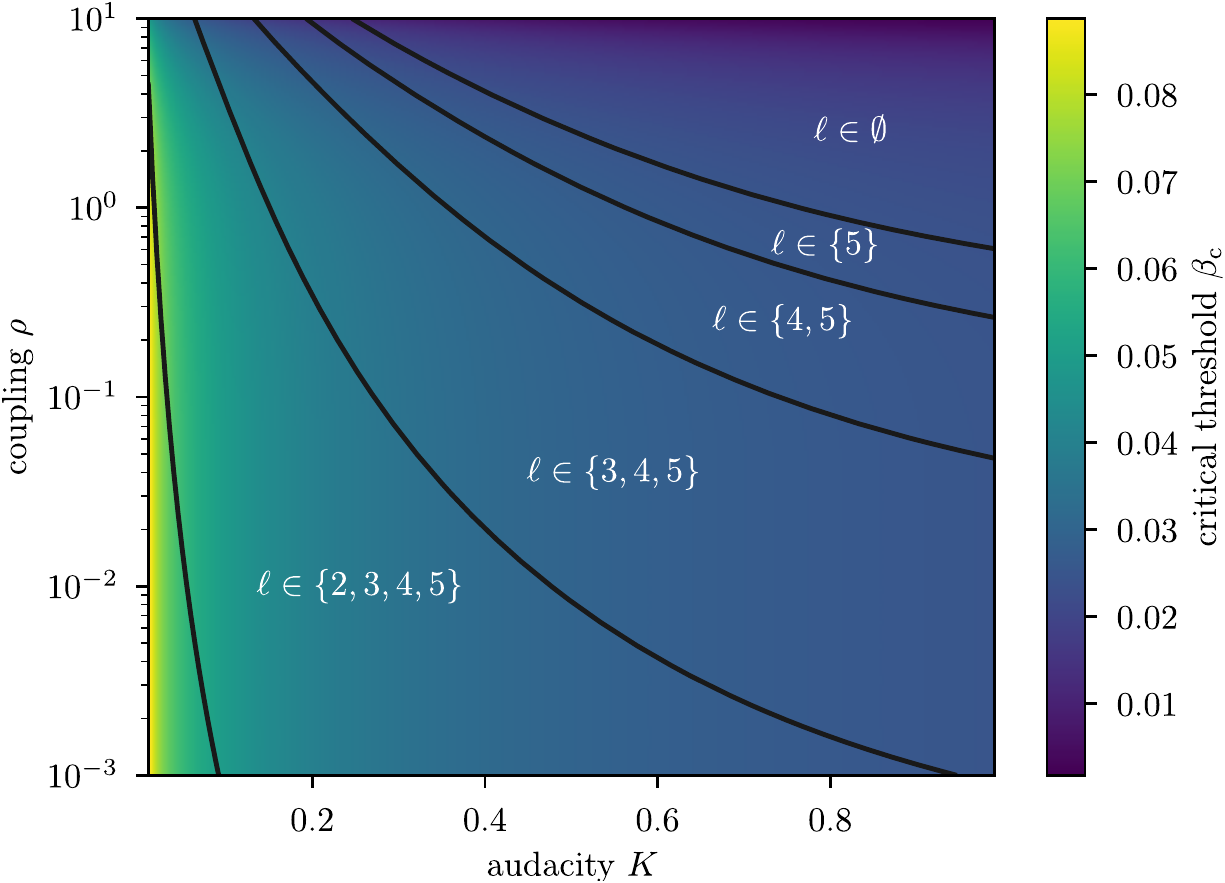}
\caption{\textbf{Localization of adoption in institutional levels near the invasion threshold.} We solve for the invasion threshold $\beta_\mathrm{c}$ using our perturbative development [Eq.~\eqref{eq:invasion_threshold}] as a function of the rate of global behavioral diffusion $\rho$ and the audacity $K = (e^{-c}+\mu)/(e^c +\mu)$ of the institutions. The solid lines correspond to $\beta_\mathrm{c} = 1/n\ell$ for $\ell \in \lbrace 1,2,3,4,5 \rbrace$, and delimit regions where a certain number of institutional level $\ell$ are \textit{active} ($\ell > 1/\beta_\mathrm{c} n$) near the invasion threshold. These levels are part of the localization set $\mathcal{L} \subseteq \lbrace 1,\dots,L \rbrace$. We use $n = 20$ and $\gamma = 1$. }
\label{fig:localization_v1}
\end{figure}

In the limit $R \to 0$, $G_{i|\ell} \to \delta_{i,0}$ (note this is true not only to order zero in $\epsilon$), consequently $Z_\ell|_{R=0} = e^{-c\ell}$. Plugging this in Eq.~\eqref{eq:Gl_gen_sol} we obtain
\begin{align}
    G_\ell|_{R = 0} = C K^\ell
\end{align}
where
\begin{align}
K = \frac{e^{-c} + \mu}{e^c + \mu} \;,
\end{align}

and

\begin{align}
    C = \frac{1-K}{1-K^{L+1}} \;.
\end{align}
We call $K$ the \textit{audacity} of the institutions, since it corresponds to ratio of the probability to increase to the next level $(\ell \to \ell + 1)$ versus the probablity to decrease to the level below ($\ell \to \ell -1$) when \textit{ignoring} the benefits.
In fact, we see that $G_\ell$ is independent of $b$ when $R \to 0$.

Finally, the invasion threshold is found by solving for $\tilde{\lambda}$ the following implicit expression:
\begin{align}
    \label{eq:invasion_threshold}
    1 = \rho n C \sum_\ell a_{\ell}^{(0)} K^\ell \;.
\end{align}
The critical transmission rate corresponds to $\beta_\mathrm{c} = n \gamma \tilde{\lambda}_\mathrm{c}$, where $\tilde{\lambda}_\mathrm{c}$ is the solution to the previous equation.
In fact, we could solve for any parameter while keeping the rest fixed, finding a critical global diffusion rate $\rho_\mathrm{c}$ for instance.

Using our expression for the invasion threshold, we are able to assess wheter certain parameters will lead to a delocalized or a localized regime, and if localized, to characterize the localization set $\mathcal{L}$.
Since $\beta_\mathrm{c}$ depends only on $K$ and $\rho$ (when $n$ and $L$ are fixed), we indicate in Fig.~\ref{fig:localization_v1} the localization set (if any) associated to different regions in the parameter space $(K,\rho)$.

Intuitively, increasing the global diffusion across institutions $\rho$ reduce the size of the localization set---only the highest level institutions are able to maintain a high adoption locally.
Interestingly, increasing audacity has the same effect, and both effect culminate in a delocalized regime at top right corner of Fig.~\ref{fig:localization_v1}.
From an outside perspective, when $\mathcal{L} = \emptyset$, the complex institutions ($\ell > 1)$ appear less relevant or important, since they achieve similar outcome but at higher costs.
High global diffusion $\rho$ in fact makes the complex institutions less relevant, since the groups communicate and influence each other effectively without them.
High audacity, however, makes the complex institutions \textit{appear} less relevant, but this is only because the lower level institutions profits from the higher level ones through the global diffusion---this is another manifestation of institutional free riding.
Globally, the invasion threshold decreases with more audacious institutions, making it easier for the adoption of innovation.

The delocalization observed in Fig.~\ref{fig:localization_v1} when increasing $\rho$ could also help explain the the emergence of institutional free riding in Fig.~\ref{fig:institutions}---in the limit of a completely delocalized regime, higher levels are unfit, decreasing the global adoption of innovation.
Hence, localization at the level of institutions appear to be a key mechanism increasing the fitness of institutions, and important for their emergence.

\end{document}